\documentclass[12pt]{article}

\usepackage{scicite}
\usepackage{graphicx}

\usepackage{times}

\newenvironment{sciabstract}{%
\begin{quote} \bf}
{\end{quote}}

\newcounter{lastnote}
\newenvironment{scilastnote}{%
\setcounter{lastnote}{\value{enumiv}}%
\addtocounter{lastnote}{+1}%
\begin{list}%
{\arabic{lastnote}.}
{\setlength{\leftmargin}{.22in}}
{\setlength{\labelsep}{.5em}}}
{\end{list}}

\title{Quantum Computation as Geometry}
\author{Michael A.~Nielsen$^{\ast}$, Mark R. Dowling,\\
Mile Gu, and Andrew~C.~Doherty\\
\\
\normalsize{School of Physical Sciences, The University of Queensland,}\\
\normalsize{Queensland 4072, Australia} \\
\\
\normalsize{$^\ast$To whom correspondence should be addressed; E-mail:  nielsen@physics.uq.edu.au}
}

\date{}

\begin{document}
\baselineskip24pt \maketitle

\begin{sciabstract}
  Quantum computers hold great promise, but it remains a challenge to
  find efficient quantum circuits that solve interesting computational
  problems.  We show that finding optimal quantum circuits is
  essentially equivalent to finding the shortest path between two
  points in a certain curved geometry.  By recasting the problem of
  finding quantum circuits as a geometric problem, we open up the
  possibility of using the mathematical techniques of Riemannian
  geometry to suggest new quantum algorithms, or to prove limitations
  on the power of quantum computers.
\end{sciabstract}

Quantum computers have the potential to efficiently solve problems
considered intractable on conventional classical computers, the most
famous example being Shor's algorithm~\cite{Shor94a} for finding the
prime factors of an integer. Despite this great promise, as yet there
is no general method for constructing good quantum algorithms, and
very little is known about the potential power (or limitations) of
quantum computers.

A quantum computation is usually described as a sequence of logical
gates, each coupling only a small number of qubits. The sequence of
gates determines a unitary evolution $U$ performed by the computer.
The difficulty of performing the computation is characterized by the
number of gates used by the algorithm, which is said to be efficient
if the number of gates required grows only polynomially with the
size of problem (e.g.  with the number of digits in the number to be
factored in the case of Shor's factoring algorithm).

We develop an alternate approach to understanding the difficulty of
implementing a unitary operation $U$.  We suppose that $U$ is
generated by some time-dependent Hamiltonian $H(t)$ according to the
Schr\"odinger equation $dU/dt = -i H U$, with the requirement that at
an appropriate final time $U(t_f)=U$. We characterize the difficulty
of the computation by imposing a cost $F(H(t))$ on the
Hamiltonian control, $H(t)$.  Following~\cite{Nielsen06b}, we choose a
cost function on $H(t)$ that defines a Riemannian geometry on the
space of unitary operations.  Finding the optimal control function
$H(t)$ for synthesizing a desired unitary $U$ then corresponds to
finding minimal geodesics of the Riemannian geometry.

We will show that the minimal geodesic distance between the identity
operation and $U$ is essentially equivalent to the number of gates
required to synthesize $U$.  This result extends the work
in~\cite{Nielsen06b}, where it was shown that the minimal distance
provides a lower bound on the number of gates required to synthesize
$U$.

The power and interest of our result is that it allows the tools of
Riemannian geometry to be applied to understand quantum computation.
In particular, we can use a powerful tool --- the calculus of
variations --- to find the geodesics of the space.  Just as in general
relativity, this calculus can be used to derive the geodesic equation,
a ``force law'' whereby the local shape of space tells us how to move
in order to follow the geodesics of the manifold.

Intuitively, our results show that the optimal way of solving any
computational problem is to ``fall freely'' along the minimal geodesic
curve connecting the identity operation to the desired operation, with
the motion determined entirely by the local ``shape'' of the space.
To appreciate how striking this is, consider that once an initial
position and velocity are set, the remainder of the geodesic is
completely determined by the geodesic equation.  This is in contrast
with the usual case in circuit design, either classical or quantum,
where being given part of an optimal circuit does not obviously assist
in the design of the rest of the circuit.  Geodesic analysis thus
offers a potentially powerful approach to the analysis of quantum
computation.  However, a caveat to this optimism is that although we
know the initial position is the identity operation, we still need to
determine the initial velocity in order to find the minimal geodesic,
and this is not in general an easy problem.

Our results can also be viewed as showing that the problem of finding
minimal quantum circuits is equivalent to a problem in geometric
control theory~\cite{Jurdjevic96a}, which has had great success in
using techniques from the calculus of variations and Riemannian
geometry to solve optimal control problems.  For example, Khaneja
\emph{et al}\cite{Khaneja01a} (c.f. also~\cite{Khaneja01b,Khaneja02a})
have used geometric techniques to analyse the minimal time cost of
synthesizing two-qubit unitary operations using a fixed two-qubit
control Hamiltonian, and fast local control.

In order to choose a cost function on the control Hamiltonian $H(t)$
we first write $H(t)$ in terms of the Pauli operator expansion $H =
\sum_\sigma' h_\sigma \sigma + \sum_\sigma'' h_\sigma \sigma$, where:
(1) in the first sum $\sigma$ ranges over all possible one- and
two-body interactions, that is all products of either one or two Pauli
matrices acting on $n$ qubits; (2) in the second sum $\sigma$ ranges
over all other tensor products of Pauli matrices and the identity; and
(3) the $h_\sigma$ are real coefficients.  We then define a measure of
the cost of applying a particular Hamiltonian during synthesis of a
desired unitary operation
\begin{equation} \label{eq:metric}
  F(H) \equiv \sqrt{\sum_\sigma' h_\sigma^2 + p^2 \sum_\sigma''
    h_\sigma^2}.
\end{equation}
The parameter $p$ is a penalty paid for applying three- and more-body
terms; later we will choose $p$ to be large, in order to suppress such
terms\cite{geometry05e}.

This definition of control cost leads us to a natural notion of
distance in $SU(2^n)$. A curve $[U]$ between the identity operation
$I$ and the desired operation $U$ is a smooth function $U : [0,t_f]
\rightarrow SU(2^n)$ such that $U(0) = I$ and $U(t_f) = U$. The length
of this curve can then be defined by the total cost of synthesizing
the Hamiltonian that generates evolution along the curve:
\begin{equation}
  d([U]) \equiv \int_0^{t_f} dt \, F(H(t)).
\end{equation}
Since $d([U])$ is invariant with respect to different
parameterizations of $[U]$\cite{geometry05a}, we can always rescale
the Hamiltonian $H(t)$ such that $F(H(t)) = 1$ and the desired unitary
$U$ is generated at time $t_f = d([U])$. From now on we assume that we
are working with such normalized curves. Finally, the distance
$d(I,U)$ between $I$ and $U$ is defined to be the minimum of $d([U])$
over all curves $[U]$ connecting $I$ and $U$.

We will show that for any family of unitaries $U$ (implicitly, $U$ is
indexed by the number of qubits, $n$) there is a quantum circuit
containing a number of gates polynomial in $d(I,U)$ that approximates
$U$ to high accuracy.  In other words, if the distance $d(I,U)$ scales
polynomially with $n$ for some family of unitary operations, then it
is possible to find a polynomial-size quantum circuit for that family
of unitary operations.  Conversely, the metric we construct also has
the property, proved in~\cite{Nielsen06b}, that up to a constant
factor the distance $d(I,U)$ is a lower bound on the number of one-
and two-qubit quantum gates required to exactly synthesize $U$.
Consequently, the distance $d(I,U)$ is a good measure of the
difficulty of implementing the operation $U$ on a quantum computer.

The function $F(H)$ specified by Eq.~\ref{eq:metric} can be thought of
as the norm associated to a (right invariant) Riemannian metric whose
metric tensor $g$ has components:
\begin{eqnarray} \label{eq:metric_components}
  g_{\sigma \tau} & = & \left\{ \begin{array}{l}
      0 \mbox{ if } \sigma \neq \tau \\
      1 \mbox{ if } \sigma = \tau \mbox{ and } \sigma \mbox{ is one- or two-body}
      \\
      p^2 \mbox{ if } \sigma = \tau \mbox{ and } \sigma \mbox{ is three- or more-body}.
      \end{array}
      \right.
\end{eqnarray}
These components are written with respect to a basis for the local
tangent space corresponding to the Pauli expansion coefficients
$h_\sigma$.  The distance $d(I,U)$ is equal to the minimal length
solution to the geodesic equation, which may be
written~\cite{Arnold98a} as $\langle dH/dt, K \rangle = i \langle H,
[H,K] \rangle$.  In this expression, $\langle \cdot, \cdot \rangle$ is
the inner product on the tangent space $su(2^n)$ defined by the metric
components of Eq.~\ref{eq:metric_components}, and $K$ is an
arbitrary operator.  For our particular choice of metric components,
this geodesic equation may be rewritten as:
\begin{eqnarray}
  p_\sigma^2 \dot h_\sigma
  & = & i \sum_\tau p_\tau^2 h_\tau \tilde h_{[\sigma,\tau]},
\end{eqnarray}
where $\tilde h_{[\sigma,\tau]} = \mbox{tr}(H[\sigma,\tau])/2^n$.  A
particular class of solutions to this equation was studied
in~\cite{Nielsen06b}, but understanding the general behaviour of the
geodesics remains a problem for future research~\cite{geometry05b}.
We note that there are powerful tools in Riemannian geometry (see,
e.g.,~\cite{Milnor69a,Berger03a}) available for the study of minimal
length geodesics.

Our goal is to use the optimal control Hamiltonian $H(t)$ to
explicitly construct a quantum circuit containing a number of gates
polynomial in $d(I,U)$, and which approximates $U$ closely. The
construction combines three main ideas, which we express through three
separate lemmas, before combining them to obtain the result
(Fig.~1).

\begin{figure}
\centering
\includegraphics[width=12cm]{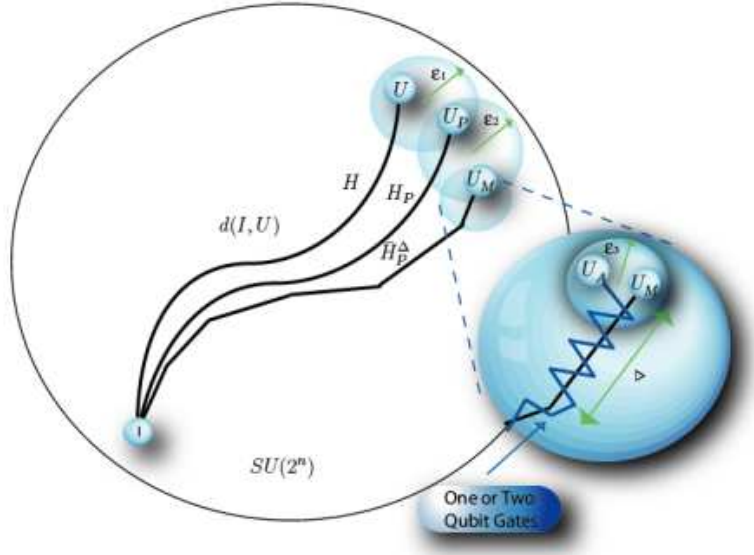}
\caption{ Schematic of the three steps used to construct a quantum
  circuit approximating the unitary operation $U$.  The circuit is of
  size polynomial in the distance $d(I,U)$ between the identity and
  $U$.  First we project the Hamiltonian $H(t)$ for the minimal
  geodesic path onto one- and two-qubit terms, giving $H_P(t)$.  By
  choosing the penalty $p$ large enough ($p=4^n$) we ensure the error
  in this approximation is small, $\epsilon_1 \leq d(I,U)/2^n$.  Next
  we break up the evolution according to $H_P(t)$ into $N$ small time
  steps of size $\Delta=d(I,U)/N$, and approximate with a constant
  mean Hamiltonian $\bar{H}^j_P$ over each step. Finally we
  approximate evolution according to the constant mean Hamiltonian
  over each step by a sequence of one- and two-qubit quantum gates.
  The total errors, $\epsilon_2$ and $\epsilon_3$, introduced by these
  approximations can be made smaller than any desired constant by
  choosing the step size $\Delta$ sufficiently small, $\Delta=O(1/(n^2
  d(I,U)))$.  In total we need $O(n^6 d(I,U)^3)$ quantum gates to
  approximate $U$ to within some constant error which can be made
  arbitrarily small.}
\label{fig:path}
\end{figure}

The first lemma shows that the error that arises by simply ignoring
the many-body interactions in $H(t)$ can be made small by choosing
the penalty $p$ appropriately. We define $H_P$ to be the
projected Hamiltonian formed by deleting all three- and
more-body terms in the Pauli expansion. Then the following result is
proved in the supporting online materials.

\textbf{Lemma~1:} Let $H_P(t)$ be the projected Hamiltonian obtained
from a Hamiltonian $H(t)$ generating a unitary $U$.  Let $U_P$ be the
corresponding unitary generated by $H_P(t)$.  Then
\begin{equation}
  \| U-U_P \| \leq \frac{2^n d([U])}{p},
\end{equation}
where $\| \cdot \|$ is the operator norm\cite{geometry05c}, and $p$ is the
penalty parameter appearing in the definition of the metric. Thus,
by choosing $p$ sufficiently large, say $p = 4^n$, we can ensure
that $\|U~-~U_P~\|~\leq~d([U])/2^n$.

Motivated by the preceding lemma, we change our aim from accurately
synthesizing $U$ to accurately synthesizing $U_P$.  To do this, we
break the evolution according to $H_P(t)$ up into many small
intervals, each of length $\Delta$.  The next lemma shows that
evolution according to the time-dependent Hamiltonian $H_P(t)$ over
such a small time interval can always be accurately simulated by a
constant mean Hamiltonian, which we denote
$\bar{H}_P^\Delta$.

\textbf{Lemma~2:} Let $U$ be an $n$-qubit unitary generated by
applying a time-dependent Hamiltonian $H(t)$ satisfying $\|H(t) \|
\leq c$ over a time interval $[0,\Delta]$.  Then defining the
mean Hamiltonian $\bar H \equiv \frac{1}{\Delta} \int_0^\Delta
dt \, H(t)$ we have:
\begin{eqnarray}
  \| U - \exp(-i\bar{H} \Delta) \| \leq 2 (e^{c \Delta}-1-c \Delta) =
  O(c^2 \Delta^2).
\end{eqnarray}

The proof of this lemma is based on the Dyson operator expansion and
is presented in the appendix.  To apply this lemma to $H_P(t)$, note
that elementary norm inequalities and the observation $F(H_P(t))
\leq 1$ imply that\cite{geometry05d} $\| H_P(t) \| \leq
\frac{3}{\sqrt{2}} n F(H_P(t)) \leq \frac{3}{\sqrt{2}} n$. Lemma~2
implies that over a time interval $\Delta$ we have:
\begin{eqnarray}
  \| U_P^\Delta - \exp(-i \bar{H}_P^\Delta \Delta) \|
  \leq 2 \left(e^{3/\sqrt{2} n\Delta}-\left(1+\frac{3}{\sqrt{2}}n\Delta
      \right)\right) = O(n^2 \Delta^2),
\end{eqnarray}
where $U_P^\Delta$ is the evolution generated by $H_P(t)$ over the
time interval $\Delta$, and $\bar H_P^\Delta$ is the corresponding
mean Hamiltonian.

Our third and final lemma shows that evolution according to a
time-independent Hamiltonian $H$ containing only one- and two-body
terms can be very accurately simulated using a number of quantum gates
that is not too large.

\textbf{Lemma~3:} Suppose $H$ is an $n$-qubit two-body Hamiltonian
whose Pauli expansion coefficients satisfy $|h_\sigma| \leq 1$. Then
there exists a unitary $U_A$, satisfying
\begin{eqnarray}
  \| e^{-iH \Delta} - U_A \| \leq c_2 n^4 \Delta^3,
\end{eqnarray}
that can be synthesized using at most $c_1 n^2 / \Delta$ one- and
two-qubit gates, where $c_1$ and $c_2$ are constants.

This result follows from standard procedures for simulating quantum
evolutions using quantum gates (see, e.g., Chapter~4
of~\cite{Nielsen00a}), and is proved in the appendix.  Note that the
average Hamiltonian $\bar{H}_P^\Delta$ provided by Lemma~2 satisfies
the assumptions of Lemma~3, since the Pauli expansion coefficients
of $H_P(t)$ satisfy $|h_\sigma| \leq 1$ for all times.

To integrate Lemmas~1-3, suppose $H(t)$ is the time-dependent
normalized Hamiltonian generating the minimal geodesic of length
$d(I,U)$.  Let $H_P(t)$ be the corresponding projected Hamiltonian,
which generates $U_P$ and satisfies $\| U - U_P \| \leq d(I,U) /
2^n$, as guaranteed by Lemma~1, and where we have chosen $p = 4^n$
as the penalty.  Now divide the time interval $[0,d(I,U)]$ up into a
large number $N$ of time intervals each of length $\Delta = d(I,U) /
N$. Let $U_P^j$ be the unitary operation generated by $H_P(t)$ over
the $j$th time interval.  Let $U_M^j$ be the unitary operation
generated by the corresponding mean Hamiltonian.  Then Lemma~2
implies that:
\begin{eqnarray}
  \| U_P^j - U_M^j \| \leq 2 (e^{3/ \sqrt{2} n\Delta} -(1+\frac{3}{\sqrt{2}}n\Delta)).
\end{eqnarray}
Lemma~3 implies that we can synthesize a unitary operation $U_A^j$
using at most $c_1 n^2/\Delta$ one- and two-qubit gates, and
satisfying $\| U_M^j - U_A^j \| \leq c_2 n^4 \Delta^3$.

Putting all these results together and applying the triangle
inequality repeatedly, we obtain:
\begin{eqnarray} \label{eq:global_1}
\| U-U_A \| & \leq & \|U-U_P\| + \| U_P - U_A \|
\\ \label{eq:global_2}
 & \leq & \frac{d(I,U)}{2^n} + \sum_{j=1}^N \| U_P^j - U_A^j \| \\
\label{eq:global_3}
 & \leq & \frac{d(I,U)}{2^n} + \sum_{j=1}^N \left( \|U_P^j-U_M^j \| + \|U_M^j-U_A^j\| \right) \\
\label{eq:global_4}
 & \leq & \frac{d(I,U)}{2^n} + 2 \frac{d(I,U)}{\Delta} \left(e^{(3/ \sqrt{2}) n\Delta} -\left(1+\frac{3}{\sqrt{2}}n\Delta\right)\right) + c_2 d(I,U)n^4 \Delta^2.
\end{eqnarray}
Provided we choose $\Delta$ to scale at most as $1/(n^{2} d(I,U))$, we
can ensure that the error in our approximation $U_A$ to $U$ is small,
while the number of gates scales as $n^6 d(I,U)^3$.  Summing up, we
have the following theorem~\cite{geometry05f}:

\textbf{Theorem:} Using $O(n^6 d(I,U)^3)$ one- and two-qubit gates
it is possible to synthesize a unitary $U_A$ satisfying $\| U -
U_A\| \leq c$, where $c$ is any constant, say $c = 1/10$.

Our results demonstrate that, up to polynomial factors, the optimal
way of generating a unitary operation is to move along the minimal
geodesic curve connecting $I$ and $U$. Since the length of such
geodesics also provides a lower bound on the minimal number of quantum
gates required to generate $U$, as shown in~\cite{Nielsen06b}, the
geometric formulation offers an alternative approach which may suggest
efficient quantum algorithms, or provide a way of proving that a given
algorithm is indeed optimal.

It would, of course, be highly desirable to completely classify the
geodesics of the metric we construct. An infinite class of such
geodesics has been constructed in~\cite{Nielsen06b}, and shown to
have an intriguing connection to the problem of finding the closest
vector in a lattice. In future, a more complete classification of
the geodesics could provide significant insight on the potential
power of quantum computation.
\\
\begin{appendix}

\noindent \textbf{Appendix:}\\
\\ \textbf{Proof of Lemma 1:}  We require three facts about the
(unitarily-invariant) operator norm $\| \cdot \|$ and the cost
function $F(H)$:

\begin{enumerate}
\item Suppose time-dependent hamiltonians $H(t)$ and $J(t)$ generate
  unitaries $U$ and $V$, respectively, according to the time-dependent
  Shr\"odinger equation.  By repeatedly applying the triangle
  inequality and the unitary invariance of $\| \cdot \|$, we obtain
  the inequality:
\begin{equation}
\| U -V\| \leq \int dt \| H(t)-J(t) \|.
\end{equation}

\item If $H$ contains only three- and more-body terms we have $F(H) =
  p \| H\|_2$, where $\| \cdot \|_2$ is the Euclidean norm with
  respect to the Pauli expansion coefficients.

\item For any $H$
\begin{eqnarray}
  \nonumber \|H \| & = & \left\| \sum_\sigma h_\sigma \sigma \right\|
  \leq \sum_\sigma |h_\sigma| \leq 2^n \| H \|_2,
\end{eqnarray}
where the final inequality follows by an application of the
Cauchy-Schwarz inequality.
\end{enumerate}

Combining these observations we have
\begin{eqnarray}
   d([U]) & = & \int dt \, F(H(t)) \\
   & \geq & \int dt \, F(H(t)-H_P(t)) \\
   & \geq & \int dt \, p \|H(t)-H_P(t) \|_2 \\
   & \geq & \frac{p}{2^n} \int dt \, \|H(t)-H_P(t) \| \\
   & \geq & \frac{p}{2^n} \| U- U_P \|,
\end{eqnarray}
from which the result follows.

\textbf{Proof of Lemma 2:} Recall the Dyson series (see,
e.g.,~\cite{Sakurai94a}, p.~325-326):
\begin{equation}
  U = \sum_{m=0}^\infty (-i)^m \int_0^\Delta dt_1 \int_0^{t_1} dt_2
  \ldots \int_0^{t_{m-1}} dt_m \, H(t_1) H(t_2) \ldots H(t_m).
\end{equation}
Note that in our finite-dimensional setting this series is always
convergent. By writing out the power series for $\exp(-i \bar H
\Delta)$ and canceling the $O(\Delta^0)$ and $O(\Delta^1)$ terms and
applying the triangle inequality we have
\begin{eqnarray}
\nonumber \| e^{-i \bar{H} \Delta} - U \| &\leq& \sum_{m=2}^\infty
\frac{\|(-i
\bar{H} \Delta)^m \|}{m!} \\
&& +  \int_0^\Delta dt_1 \int_0^{t_1} dt_2 \cdots \int_0^{t_{m-1}} dt_m \| H(t_1) H(t_2) \cdots H(t_m) \| \\
&\leq& 2 \sum_{m=2}^\infty \frac{c^m \Delta^m}{m!} = 2 \left( e^{c
\Delta}-1-c \Delta \right),
\end{eqnarray}
where for the second line we have used the standard norm inequality
$\| X  Y \| \leq \| X \| \| Y \|$, the condition $\|H(t)\| \leq c$
and the fact that $\int_0^\Delta dt_1 \int_0^{t_1} dt_2 \cdots
\int_0^{t_{m-1}} dt_m = \Delta^m / m!$. This is the required result.
\textbf{QED}

\textbf{Proof of Lemma 3:}  We use standard quantum simulation
techniques, see e.g.\ Section~4.7 of~\cite{Nielsen00a}.  Divide the
interval $[0,\Delta]$ up into $N=1/\Delta$ steps of size $\Delta^2$.
Define
\begin{equation}
\label{eq:deltatgate} U_{\Delta^2} = e^{-i h_1 \sigma_1 \Delta^2}
e^{-i h_2 \sigma_2 \Delta^2} \ldots e^{-i h_L \sigma_L \Delta^2}
\end{equation}
where $L=O(n^2)$ is the number of terms (all one- and two-body) in
$H$.  Each factor on the right hand side of this equation is an
allowed one- or two-qubit quantum gate.  Using, e.g., Eq.~(4.103) on
page~208 of~\cite{Nielsen00a} it is straightforward to show that:
\begin{equation}
\label{eq:deltatgateerror } U_{\Delta^2} = e^{-i H \Delta^2} + O(L^2
\Delta^4),
\end{equation}
and therefore through repeated applications of the triangle
inequality and the unitary invariance of the operator norm we
obtain:
\begin{equation}
\label{eq:totalerror} \|e^{-i H \Delta} - U_{\Delta^2}^N \| \leq c_2
N n^4 \Delta^4 = c_2 n^4 \Delta^3,
\end{equation}
where $c_2$ is a constant.  Eq.~(\ref{eq:totalerror}) shows how to
approximate $e^{-i H \Delta}$ using at most $c_1 n^2 / \Delta$
quantum gates, where $c_1$ is another constant, which is the desired
result. \textbf{QED}

\end{appendix}

\newpage

\begin{thebibliography}{10}
\bibitem{geometry05e}
An alternative way of viewing this cost function is as a penalty metric of the
  kind used in sub-Riemannian geometry. See, for example, p.~18
  of~\cite{Montgomery02a}.
\bibitem{geometry05a}
The cost function has the property that $F(\alpha H) = |\alpha| F(H)$; this
  property and the chain rule imply invariance of the length with respect to
  reparameterization.
\bibitem{geometry05b}
Our metric is superficially similar to the usual metric of Euclidean space, and
  it is tempting to suppose that geodesics must be ``straight lines", i.e.\
  constant Hamiltonians. However the Pauli coefficients for the Hamiltonian
  actually correspond to a changing local basis for the tangent space, not a
  fixed basis, and hence constant Hamiltonians do not, in general, give rise to
  geodesics.
\bibitem{geometry05c}
The operator norm of $X$ is defined as $\|X\| = \max_{|\psi \rangle} |\langle
  \psi | X |\psi \rangle |$, where the maximization is over all normalized
  vectors, $|\langle \psi | \psi \rangle|^2 = 1$.
\bibitem{geometry05d}
The first inequality comes from the fact that there are $9n(n-1)/2 + 3n$ one-
  and two-qubit terms.
\bibitem{geometry05f}
Note that the overhead factors in this theorem may be substantially improved,
  e.g., by making use of higher-order analyses in lemmas~1-3. However, the key
  point --- that $U$ can be accurately approximated with a number of gates that
  scales polynomially with $d(I,U)$ --- remains the same.
\end{thebibliography}

\begin{scilastnote}
\item Thanks to Scott Aaronson, Mark de Burgh, Jennifer Dodd, Charles
  Hill, Andrew Hines, Austin Lund, Lyle Noakes, Mohan Sarovar, and Ben
  Toner for helpful discussions and the Australian Research Council
  for funding.  We are especially grateful to Lyle Noakes for pointing
  out to us the simplified form of the geodesic equation for
  right-invariant metrics.
\end{scilastnote}

\end{document}